%% file: ms.tex
\documentclass[acmsmall]{acmart}

\usepackage{makecell}
\usepackage{rotating}
\usepackage{multirow}
\usepackage{makecell}
\usepackage{MCZ-traffic-survey-2019}
\usepackage[utf8]{inputenc}

\begin{document}

\title[Coordination of Autonomous Vehicles: Taxonomy and Survey]{Coordination of Autonomous Vehicles:\\ Taxonomy and Survey}

\author{Stefano Mariani}
\email{stefano.mariani@unimore.it}
\orcid{0000-0001-8921-8150}
\author{Giacomo Cabri}
\email{giacomo.cabri@unimore.it}
\orcid{0000-0002-4942-2453}
\author{Franco Zambonelli}
\email{franco.zambonelli@unimore.it}
\orcid{0000-0002-6837-8806}
\affiliation{%
  \institution{Universit\`a degli Studi di Modena e Reggio Emilia}
}

\renewcommand{\shortauthors}{Mariani, Cabri, Zambonelli}

\begin{abstract}
In the near future, our streets will be populated by myriads of autonomous self-driving vehicles to serve our diverse mobility needs. This will raise the need to coordinate their movements in order to properly handle both access to shared resources (e.g., intersections and parking slots) and the execution of mobility tasks (e.g., platooning and ramp merging). In this paper, we firstly introduce the general issues associated to coordination of autonomous vehicles, by identifying and framing the key classes of coordination problems. Following, we overview the different approaches that can be adopted to manage such coordination problems, by classifying them in terms of the degree of autonomy in decision making that is left to autonomous vehicles during coordination. Finally, we overview some further peculiar challenges that research will have to address before autonomously coordinated vehicles can safely hit our streets.
\end{abstract}



\maketitle

\section{Introduction}\label{sec:introduction}

Autonomous self-driving vehicles will soon populate our streets~\cite{preparing,coppola2016}. Yet already several car models include at least some level of driving autonomy (e.g., automated steering and lane-keeping, or emergency stop in the presence of obstacles), but still definitely require the driver to be in control for some, if not most, of the time. However, it won't be long before increasing levels of automation will make the driver mostly, if not fully, unnecessary.

Besides the advantage of relieving us from the duty of driving and paying attention, thus making it possible to exploit travel time in other activities, autonomous vehicles will bring further important benefits. They will reduce crashes, now mostly due to bad human behaviours and human errors, most likely saving millions of injuries and lives~\cite{bimbraw2015}. They will notably reduce the number of circulating vehicles and, also thanks to route optimisations, will definitely reduce traffic and pollution~\cite{menon2018}. Last but not least, they will pave the way for a number of innovative solutions in the provisioning of mobility services~\cite{roth2018}, to serve our needs with much greater levels of quality and efficiency than today. 

Innovative mobility solutions that will be promoted (or boosted) by autonomous vehicles include, just to mention a few: car sharing~\cite{hao2018}, where fleets of autonomous vehicles (whether provided by public actors or by private companies) will be available to serve our urban mobility needs; personalised public transport and ride sharing ~\cite{stiglic2018,Bicocchi17}, where autonomous vehicles and buses can dynamically gather people based on their actual required routes; smart and more effective parking approaches~\cite{7895130}, in that autonomous vehicles can search for parking slots based on criteria different from the ``very soon and very close'' one that we human usually adopt, and exploiting additional information that they might have.   

Most of the current industrial and applied research in the area of autonomous vehicles concerns the methods and tools to enable \emph{individual} autonomous vehicles to hit the road safely. However, it is getting increasingly recognised that, to get full advantage from autonomous vehicles (enabling overall traffic reduction and pollution, and supporting the effective implementation of the above mentioned mobility solutions), a number of situations will compulsory require \emph{coordinating} the relative activities and movements of vehicles~\cite{SassiZ14,AbeywickramaMZ18}.
This trend is witnessed by initiatives such as the Grand Cooperative Driving Challenge, which in its latest edition explicitly focussed on \emph{cooperative automated driving}~\cite{7553038}.

Examples of very diverse situations that require a proper and careful coordination amongst groups of vehicles include: crossing intersections, entering a motorway, platooning, organising urban deployment and rides for fleets of ride/car-sharing vehicles, trying to improve parking occupancies and reduce parking times. Effectively supporting such coordination needs implies devising effective mechanisms and strategies to support coordination activities. 

\subsection{Contribution}

Against this background, the overarching contribution of this paper is to overview and frame the key issues associated to the coordination of autonomous vehicles, and overview the possible approaches to attack the problem. To this end, we:

\begin{itemize}

	\item Introduce a taxonomy to classify the key problems requiring some forms of coordination, and, accordingly, present and discuss the main issues associated with approaching the problems. In particular, we show that vehicles coordination problems can be classified in terms of \emph{resource-oriented} vs.\ \emph{task-oriented} and, orthogonally, in terms of \emph{competitive} vs.\ \emph{cooperative} ones, and that the issues associated to them are mostly analogues to those traditionally expressed in \emph{coordination}~\cite{coordination-csur26,selforgcoord-ker25years,techsurvey-coord2018} and \emph{concurrency} theory~\cite{Coffman:1971:SD:356586.356588,Hoare:1978:CSP:359576.359585,Dijkstra2002}.
	
	\item Identify the different approaches, e.g., coordination mechanisms, protocols, and strategies, that can be put at work to solve the identified issues. In particular, we show that (independently of the specific coordination problem to address) the different approaches can be classified in terms of the \emph{degree of autonomy in decision making} that is left to individual vehicles during the coordination process---not to be confused with the ``level of autonomy'' in driving defined in~\cite{SAE2018}.

	\item Overview some research challenges that horizontally apply to most (if not all) the identified coordination problems, and that call for additional inter-disciplinary research to identify solutions that are both technically sound and socially acceptable.
	
\end{itemize}
To the best of our knowledge, this is the first survey investigating the \emph{problems} associated to vehicle coordination in general broad terms. Other surveys exist but focus on specific coordination issues (e.g., either intersection management~\cite{Rios-Torres20171066} or smart parking~\cite{7895130}) and are mostly concerned with computational techniques and infrastructural requirements of \emph{solutions}, rather than on coordination problems. Also, again to the best of our knowledge, this is the first attempt at providing a comprehensive taxonomy of coordination problems for autonomous vehicles and at surveying under a unifying perspective the spectrum of possible solutions. 

\subsection{Scope}

Concerning the extent of applicability of our analysis with respect to the level of autonomy of vehicles (as per the autonomy level classification defined in~\cite{SAE2018}), it applies to level 5 (fully autonomous vehicles requiring no drivers) as well as to level 4 (vehicles that can drive in autonomy in the vast majority of situations, but that can still reclaim the human driver in specific contingencies).

We also emphasise that this article is not a systematic literature review. From the literature, we have tried to identify and discuss what we consider key proposals in the area, in order to give readers a sound overview of problems and solutions, each properly supported by relevant examples\footnote{Yet, we would really appreciate if the reviewers could suggest us references to relevant articles that we might have miss.}.  

 \subsection{Structure}
 
The remainder of this paper is structured as follows: Section~\ref{taxonomy} introduces the taxonomy of coordination problems; Section~\ref{survey} classifies and overviews the approaches that can be adopted to address the aforementioned problems; Section~\ref{challenges} introduces some further coordination challenges that are horizontal w.r.t.\ the taxonomy and the approaches; Section~\ref{conclusions} concludes the paper.

\section{Taxonomy of Coordination Problems}\label{taxonomy}

The problem of \emph{coordinating} the movements and activities of autonomous, connected vehicles in an urban environment is a very general one, including highly heterogeneous scenarios from intersection management to car sharing.
Nevertheless, such heterogeneous scenarios can all be homogeneously framed in a single conceptual framework. 

\subsection{Definitions}\label{ssec:defs}

With ``coordination'' we refer to the \emph{decision making} process (involving vehicles themselves and possibly some additional infrastructural process) aimed at \emph{orchestrating} vehicles' actions so as to achieve a goal which cannot be achieved (or not optimally) by each vehicle in \emph{isolation}. The goal may belong either to individual vehicles (e.g., crossing an intersection), or to a group of vehicles (e.g., allocating parking slots to the vehicles of a company fleet), or to the traffic system as a whole (e.g., balance overall traffic distribution in a city). 

In particular, we can define a two-dimensional taxonomy for the major classes of coordination problems---depicted in \xt{problems}. 
On one dimension, the problems can be categorised according to whether they are \emph{resource-oriented} or \emph{task-oriented}:

\begin{itemize}

\item A resource-oriented coordination problem concerns enabling vehicles to acquire access to a shared limited resource (e.g., a shared intersection), and requires the coordination process to safely regulate such access according to specific strategies and rules.

\item A task-oriented coordination problem concerns enabling vehicles to complete a specific task (e.g., bring a group of persons home), and requires the coordination process to properly allocate to vehicles the responsibilities and actions required to achieve such task.
 
\end{itemize}
It is worth noting that a resource-oriented coordination process always takes place in the context of a task to be executed (e.g., ``I need to cross an intersection to bring a person home'') and that, viceversa, completing the execution of an allocated task may require access to some shared resource and the involvement in further coordination processes. 
On the other dimension, coordination problems can be of a \emph{collaborative} nature or of a \emph{competitive} one:

\begin{itemize}

\item A collaborative coordination problem involves vehicles that either share a common overall goal (e.g., the vehicles of a company fleet) or that recognise that cooperating is the best way to achieve their own individual goals (e.g., group of trucks platooning in a motorway).

\item A competitive coordination problem involves self-interested vehicles that have the primary intent to achieve their own goal and have no interest for cooperating with others (e.g., crossing an intersection or finding a parking slot).

\end{itemize}
%
Regardless of the specific bi-dimensional classification of the coordination problem at hand, the coordination process aimed at solving the problem has to satisfy the following general properties (assuming a specific form depending on the given problem, as depicted in \xt{mapping}):

\begin{itemize}

\item \emph{Safety}, expressing that ``something bad \emph{never} happens'' during the coordination process (e.g., two cars never crash while crossing an intersection)---where ``never'' may be possibly replaced by a bounded probability.
\item \emph{Liveness}, expressing that ``something good \emph{eventually} happens'' (e.g., all cars will eventually manage to cross an intersection)---where ``eventually'' may be possibly replaced by a bounded time horizon.

\end{itemize}
In addition, all coordination problems can be associated to some specific \emph{Quality} measure, expressing how well a coordination process manages to solve the coordination problem (e.g., the average or cumulative delay at which cars manage to cross an intersection).

\subsection{Problems statement}
Let us now analyse in detail, according to the concepts introduced in the previous subsection, the key coordination problems that will arise in future autonomous vehicles scenarios. These include:
\begin{itemize}
	\item crossing intersections; 
	\item parking, in the case of both private vehicles and public (or company) fleets; 
	\item ride sharing (there including carpooling), again in the case of individual vehicles or fleets; 
	\item ramp merging (or the similar lane changing and roundabout crossing problems); 
	\item platooning;
	\item traffic flow optimisation.
\end{itemize}
In \xt{mapping} we categorise these problems according to our proposed taxonomy, and further detailing their peculiar features from a coordination perspective. The following subsections thoroughly describe each problem, why it has been categorised in the given way, and how it can be framed as a typical coordination problem as stemming from classical coordination and concurrency theory.
%
\input{problems}

\subsubsection{Intersection management}

Intersection management is the problem of coordinating vehicles while concurrently crossing intersections~\cite{Rios-Torres20171066}. As such, it is a \emph{competitive}, \emph{resource-oriented} problem, since vehicles are self-interested agents willing to obtain the right-of-way (as soon as possible) across the shared resource represented by the intersection.
As for the properties, intersection management should enable vehicles to \emph{safely} cross an intersection by avoiding collisions, eventually giving each vehicle the right-of-way (\emph{liveness}), while possibly minimising the average \emph{delay} experienced by vehicles in waiting the right-of-way (\emph{quality measure}). 

Today, intersection management is realised either by a central controller, the traffic light, or by imposing to vehicles (i.e., to their drivers) pre-defined coordination rules to be obeyed (e.g., stop at sign or give right-of-way to vehicles coming from the right). This puts the responsibility for safety fully in charge of humans, and does not promote efficiency. Future autonomous vehicles scenarios will make it possible to conceive a variety of automated solutions, safer and more efficient, eventually making traffic lights and stop signs obsolete.  

\input{mapping}

\subsubsection{Smart parking}\label{sssec:parking}

Smart parking is the problem of coordinating vehicles, either privately owned or belonging to a fleet, while looking for parking slots~\cite{Polycarpou2013,7895130}. As such, it is a \emph{resource-oriented} problem that, in the case of privately owned vehicles (as it is most often the case today), assumes a \emph{competitive} form, since free parking slots in cities are typically scarce resources very hard to find, and private vehicles have no interest in collaborating with each other to fairly share parking slots. 

However, as cities are also getting (and will be increasingly) populated by ``fleets'' of vehicles made available by private companies or municipalities, for short term renting or car sharing~\cite{Hoch15}, vehicles belonging to the same fleet can indeed have interest in fairly allocating the available parking slot to the fleet members, and in \emph{cooperating} towards such a goal.  

As for the properties, coordinating vehicles while they are seeking for free parking implies for safety avoiding overbooking (allocating the same parking slot to more than one vehicle), and for liveness avoiding starvation (i.e., vehicles that gets never assigned to a parking slot). Quality metrics can be defined according to a variety of parameters (i.e., quickness in finding a slot, but also distance of the slot from the desired destination, and hourly price for occupancy), and different vehicles or different fleets can each have their own personal quality metrics (e.g., an elderly people may prefer taking a little more time to park in order to find a slot closer to destination).  

Today, in most cities worldwide, parking slots (and the movement of vehicles wandering in their search) are simply not managed. Some cities provide digital signals to inform human drivers about availability of parking slots in specific parking areas, in order to direct vehicles towards them. A few cities provide mobile apps for discovering in real-time free parking slots and possibly book them~\cite{5928901,s141222372}. In the presence of autonomous connected vehicles, though, it will be easy to enhance such embryonic parking management approaches towards much higher levels of efficiency and quality, without having humans to care about.  

\subsubsection{Ride sharing}

Ride sharing is the problem of coordinating groups of vehicles to collectively satisfy a number of ``mobility requests'', each concerned with bringing a person from its current location to a desired destination~\cite{Bicocchi17}. However, the same coordination problem can be instantiated in many other different application domains, for instance to merchandise delivery. 

As such, ride sharing is a \emph{task-oriented} problem: given a number of mobility requests expressed at a given time (the tasks to be performed) and a number of vehicles currently available to serve them, the system must assign vehicles to requests. Ride sharing takes the form of a \emph{competitive} problem in the case of privately owned vehicles. Indeed, it is expected that in the future, privately owned autonomous vehicles, when not in use by owners, can be instructed to act as ride sharing vehicles to monetise otherwise idle time. And clearly, such private vehicles will compete with each other for being assigned a mobility request to satisfy, and maximise their personal gains. However, in the presence of fleets of ride sharing vehicles, the problem becomes a \emph{collaborative} one, in that the vehicles of the fleet have interest in coordinating with each other to maximise the overall gain of the fleet, and the overall satisfaction of users. Indeed, this is what already happens, to some extent, with the fleets of taxi companies. 

As for the properties, safety concerns not assigning the same mobility request to multiple vehicles and, viceversa, not assigning multiple users with different mobility needs to the same vehicle. Liveness concerns avoiding starvation for both vehicles and users. Quality measures strongly depend on the perspective: for vehicles, it may be having coordination schemes that maximise car usage and monetary gains, whereas for users it may be having coordination schemes that minimise waiting time and limit the walking distance needed to reach the planned pick-up location~\cite{Agatz2012295}. 

It is worth emphasising here that ride sharing, as we have presented it, includes the case of carpooling: a vehicle currently moving (or planning to move) towards a destination to serve a specific user, that can become available to carry additional users that happen to share the same mobility need (i.e., are on the path of the vehicle and need to reach a destination on that path). 

Currently, coordination for ride sharing is mostly promoted at the level of municipal taxi companies or of companies like Uber and Lyft, and based on centralised policies for the assignment of tasks to vehicles that strongly rely on the availability and personal preferences of human drivers, thus limiting efficiency and user satisfaction. Autonomous vehicles will make it possible to conceive more flexible and reliable policies.

\subsubsection{Ramp merging}\label{sssec:merging}

Ramp merging is the problem of coordinating vehicles while entering and leaving highway ramps.  Lane changing is the similar problem (that is, equivalent from the coordination viewpoint) of changing lane in a multi-lane road (usually, a highway)~\cite{Rios-Torres20171066}. In addition, we emphasise that roundabout crossing is a coordination problem that can be stated as a sequence of ramp merging, lane changing, and then splitting problems.

All of the above problems are better understood as \emph{task-oriented} ones, with the key problem of efficiently steering vehicles during merging operations. Of course, the problem also somehow concerns access to a shared resource (the few meters of the lane that the merging vehicle needs to occupy). However, in normal traffic conditions, such resource in not scarce, and entering a lane from a ramp (or entering a roundabout, or changing a lane) is not focussed on that, but rather on making sure that the movements of all vehicles in the surroundings are properly orchestrated to avoid crashes and enabling the change. With this regard, although the problem is apparently competitive (the vehicles already in a lane do not care about the need of a car to enter that lane), coordination of vehicles during the process necessarily becomes \emph{cooperative}, as all vehicles have the common interest in not crashing with each other\footnote{Indeed, we emphasise that the very fact of avoiding intersections in highways, and of substituting intersections with roundabouts wherever possible in cities, is motivated by the possibility of turning a highly competitive resource-oriented problem into a more cooperative task-oriented one.}. Only in very intense traffic conditions the problem tends to become a resource-oriented and competitive one.  

As for most of the other problems, the main safety and liveness properties to be preserved in ramp merging (and equivalent) problems are to avoid collisions and starvation, respectively. Quality clearly maps to minimising the time required to perform the task.

Today, several commercial vehicles already integrate the capability to perform lane merging and changing in autonomy, but without any form of direct coordination with other vehicles. Thus, safety is still in the hand of other (human driven) vehicles, and the time needed to complete the merging procedure in their willingness and quickness to cooperate. 
 
\subsubsection{Platooning}

Platooning is the problem of coordinating manoeuvers of a fleet of vehicles so that they travel altogether as a single entity, for instance by keeping the same speed and relative positions in a highway~\cite{7056505}, as it happens in natural systems for flocking or schooling behaviours~\cite{toner1998flocks}. Platooning is a coordination problem \emph{per se}, although it is also often used as a mechanism in support of (or to facilitate attacking) other coordination problems, such as intersection management (by letting vehicles with the right-of-way to cross in platoons), traffic flow optimisation~\cite{4621202} (by avoiding that abrupt changes in speed can lead to traffic jams), and ramp merging (or lane changing)---the latter being often interpreted, in fact, as entering/leaving a platoon.

Platooning is clearly a \emph{task-oriented} coordination problem (the task being to move as a single entity) in which vehicles have incentives to cooperate (i.e., to reduce fuel consumption and risk of accidents).
As for most of the other classes of coordination problems, safety concerns are about avoiding collisions, whereas liveness and quality refer to the capability of preserving the structure of the platoon despite contingencies, and doing so with the highest adherence to the ideal shape, respectively.

Although platoons tend to naturally form in streets and highways, the explicit deliberate choice of moving in platoons is not currently enforced by regulations or ad-hoc coordination mechanisms between vehicles~\cite{7497531}, as we instead expect will happen in the future.

\subsubsection{Traffic flow optimisation}
\label{flow}

Traffic flow optimisation is the problem of coordinating vehicles' journeys across the transport network so as to achieve a maximal (or a satisfying) balancing of road infrastructure exploitation and avoid (or limit) traffic congestions and jams~\cite{6083055,5978226}. Clearly, this is a large-scale general coordination problem, that also somewhat subsume all the previously presented problems. In fact, traffic flow optimisation implies also finding good solutions to cross intersections efficiently, finding parkings without having cars wander around hopelessly, exploiting ride-sharing as much as possible, and quickly merging into highway lanes, since all these measures  improve utilisation of the road infrastructure---hence have a positive impact on overall traffic congestion. However, the objective of traffic flow optimisation should also integrate large-scale strategies to route vehicles across the road network in order to promote load balancing of the network itself (and of course without forgetting where each individual vehicle needs to go).

Such a coordination problem is clearly a \emph{resource-oriented} one, the resource to be allocated being the road network. At first, the problem may appear of a \emph{competitive} nature, since individual vehicles have the selfish goal of reaching destination fast, regardless of others. However, it could naturally become \emph{cooperative} if, as it should be, vehicles are made aware that routing suggestions (or the routing plans that they elaborate together) can facilitate all individual vehicles in reaching their destinations smoothly.

In traffic flow optimisation, safety amounts at avoiding congestions and traffic jams, while liveness amounts at routing vehicles so as to avoid loops and never-ending trips. In terms of quality measure, the degree of balance in the exploitation of the road network and the overall fluidity of the traffic flow are certainly the primary ones. However, also the extent to which the coordination process deviates vehicles from their original route should be accounted for.

Current measures adopted by municipalities to optimise traffic flow are mostly based on static strategies such as imposing one-way streets to route vehicles across specific paths or limiting access to certain zones/streets to specific categories of vehicles. However, such measures cannot cope with the dynamics of traffic conditions and are largely inefficient in handling the diverse situations that happen at different hours of the day or periods of the year~\cite{6482260}. Adaptive traffic light synchronisation can work well to improve traffic flow in limited parts of a city, but cannot help at a large (e.g., urban) scale~\cite{6338817,6502719}. Digital signals adaptively suggesting directions to drivers can somewhat work, but are inherently limited by the fact that human drivers often disregard them to follow their originally planned route (or the one suggested by traffic navigator). Fortunately, such situation will change when autonomous vehicles, not drivers, will receive suggestions or cooperatively elaborate routes.

\section{Coordination Approaches}\label{survey}

In this section we overview the relevant approaches proposed so far in the literature (as well as those that can be, at least conceptually, conceived) to address  the identified vehicles coordination problems. Independently of the specific coordination problem addressed, such approaches can be classified in terms of the \emph{degree of autonomy in decision making} left to vehicles during the coordination process.

\subsection{Definitions}

By ``degree of autonomy in decision making'' we refer to the extent to which vehicles can decide their own course of actions by themselves while coordinating. Such decision of course can be based on information acquired by vehicles about the current state of the affairs, information that can be obtained by the vehicles' own sensors, by road-side infrastructural elements (vehicle-to-infrastructure, or V2I, communications) or from the other vehicles participating the coordination process (vehicle-to-vehicle, or V2V, communications).

\begin{figure}[!t]
	\centering
	\includegraphics[width=0.7\textwidth]{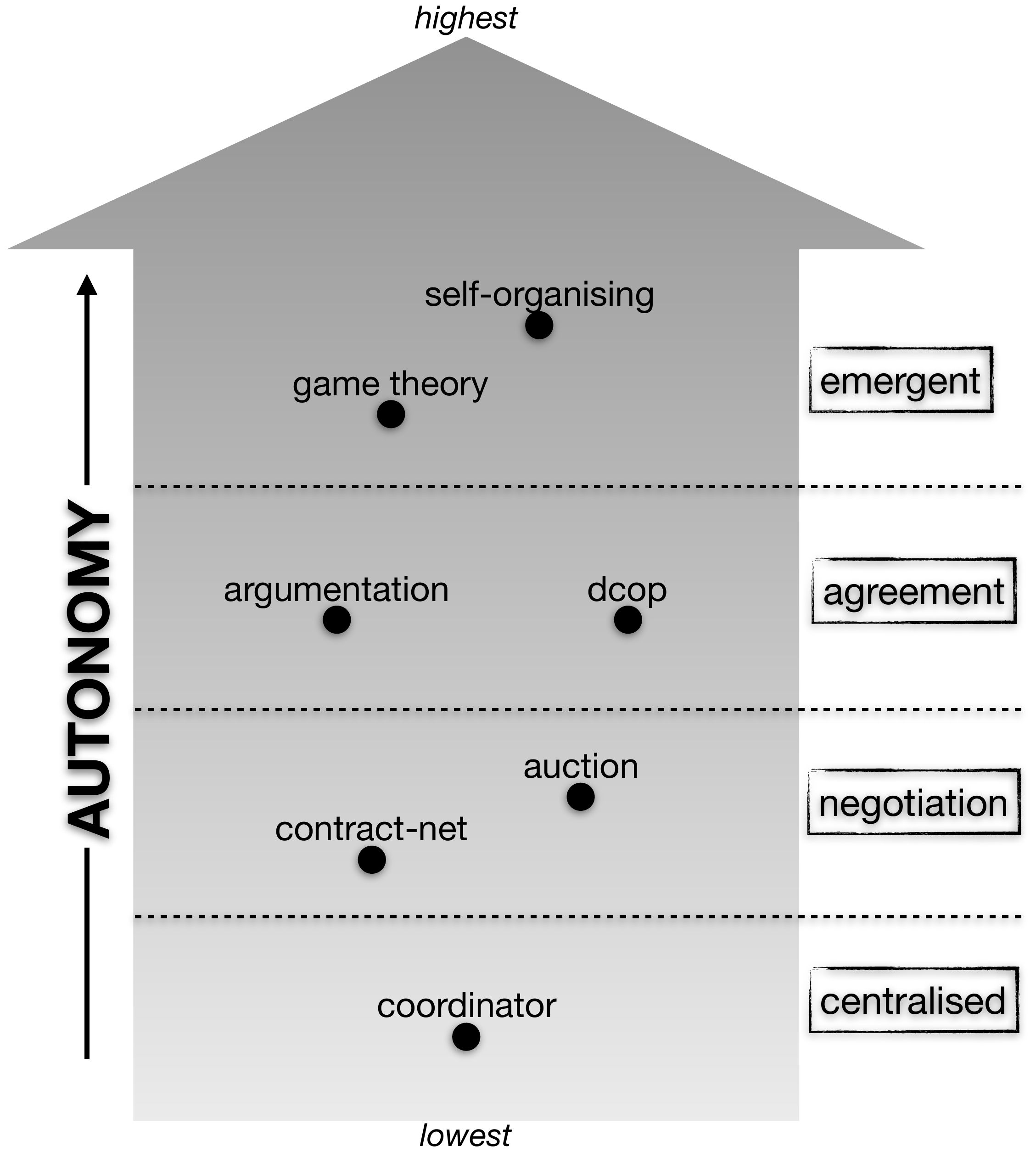}
	\caption{Coordination approaches categorised according to increasing degree of autonomy in decision making left to vehicles during the coordination process.}
	\label{fig:autonomy}
\end{figure}

By definition, in any coordination process, the entities to be coordinated cannot act completely free, and must undertake actions that account for the actions of the other entities involved in the process~\cite{GelernterC92}. Thus, there is never full autonomy and freedom. However, different approaches to coordination may leave to entities different degrees of freedom in their decision making, i.e., in selecting the actions to perform during the process. Hence, the degree of autonomy may range from fully externally imposed actions (lowest autonomy) to fully self-determined action (highest autonomy).
In particular, as depicted in Figure~\ref{fig:autonomy}, we can identify four main classes of coordination approaches centred around the concept of degree of autonomy:

\begin{itemize}
	\item \emph{Centralised}, where the burden of coordination (that is, the decision making determining the outcome of the coordination process) is entirely charged upon an individual computational entity (i.e., a \emph{coordinator}) whose decisions on how everyone should act are undebatable, and to whom vehicles must abide by design without any autonomous decision making left to them. A traditional traffic light exemplifies the role of such centralised coordinator. We emphasise that the term ``centralised'' here refers to the decision making process, not to the actual computing infrastructure supporting it, which can include for instance distributed processing of information by multiple sensors/cameras and/or services to perform reasoning in the cloud.
	\item \emph{Negotiation}, where the burden of coordination is distributed amongst the ensemble of coordinating vehicles, who participate to a specific \emph{negotiation protocol}, typically inspired by economic mechanisms. In a negotiation protocol, the vehicles involved can ``propose'' (each according to its own internal strategy and its own situation and goals) solutions and actions, amongst a set of admissible moves dictated by the protocol at each step. If properly designed, the protocol will eventually guarantee the convergence towards an equilibrium solution, determining \emph{who} (i.e., which vehicle) should do \emph{what}, and \emph{when}, to solve the coordination problem. Most representative negotiation protocols are: \emph{Contract Net}, for collaborative problems, and \emph{auctions}, for competitive ones~\cite{lopes2008negotiation}.  
	
	\item \emph{Agreement}, where coordinating vehicles participate to a \emph{dynamic protocol} defined by themselves in a collective way, in a sort of ``meta-coordination'' process whose outcome is both the set of admissible moves, now jointly defined, and possibly even a dynamic re-determination of the goals to be achieved during the coordination process. The distinguishing feature here iss the ability of agents to collectively define the protocol itself, that is, the goal to pursue and their strategy to make moves. Examples of these dynamic protocols include those based on \emph{argumentation}~\cite{rahwan2003argumentation}, where involved entities discuss and argue together to reach a common perspective on situations, goals, and solutions, and \emph{distributed constraint optimisation}~\cite{maheswaran2004taking}, where agents tries to collectively find a solution to a constraint optimisation problem.

	\item \emph{Emergent}, where coordinating vehicles do not explicitly engage in any coordination protocol, thus do not even share the goal of reaching a common agreement. Rather, every vehicle behaves in a selfish way according to its goals and to maximise utility of actions w.r.t.\ the goals, and according to the perceptions it collects about other participants to the coordination process. Examples include: \emph{game theoretic} approaches~\cite{nisan2007algorithmic}, where explicit communication is lacking, each vehicle merely assumes rationality of others, and computes its own course of actions based on informed guesses about others' expected behaviour; and \emph{self-organising algorithms}~\cite{mamei2006case}, typically nature-inspired, where vehicles act in a purely reactive way, based on the implicitly perceived presence and situation of other vehicles, typically expressed via ``traces'' (e.g., virtual pheromones or virtual computational fields) in the environment. 
	
	\end{itemize}
\input{approaches}
The criteria defining the above described categories are summarised in \xt{approaches}.
%

Let us now analyse, for each of the identified coordination problems, how different solutions based on different autonomy classes can be (or cannot be) applied, also with reference to representative literature examples (\xt{survey}).

\input{survey.tex}

\subsection{Intersection management}

Intersection management is a competitive resource-oriented coordination problem: vehicles compete for the right to acquire exclusive access to the portions of the intersection they have to cross. The goal of the coordination process is to minimise delay in handling concurrent access to the intersection, and to guarantee collision avoidance and starvation of vehicles.

\subsubsection{Centralised} 

In this category lie all the approaches in which a computational central authority (let's call it the intersection manager) bears alone the burden of decision making. The intersection manager is typically in charge of: \emph{(a)} receiving information from vehicles approaching the intersection (i.e., origin, destination, speed); \emph{(b)} elaborating a set of collision free trajectories enabling vehicles to safely cross the intersection (which may require some vehicles to slow down or change lane) \emph{(c)} precisely instructing (or directly commanding) the vehicles about what to do, or informing them about what constraints they must abide to while crossing the intersection.

It is worth emphasising that for centralised approaches to intersection management (and in general for all centralised approaches to coordination) the distinction between cooperative and competitive approaches vanishes, since the central authority handles goals, strategies, and deals with conflicts by itself.

Examples of centralised proposal to intersection management include~\cite{Wu201565}, which attack the problem in terms of a traditional mutual exclusion approach, and~\cite{6121907}, in which the authors propose a control algorithm implementing a nonlinear constrained optimisation in charge of computing the best moves for every vehicle and then directly manipulating vehicles' driving parameters. A similar stance is taken in~\cite{6338827}, where cooperative adaptive cruise control is exploited for intersection management, by assuming that a smart controller device placed in the intersection can communicate with incoming vehicles to instruct them about the actions to perform.

Other approaches are a little more permissive and let the inbound vehicles decide how to fulfil a set of constraints set by the intersection manager (which may regard the time slot assigned for crossing) as in the work by Dresner and Stone~\cite{Dresner2008591}: the authors propose a \emph{reservation-based approach} in which incoming vehicles request assignment of a time slot for crossing to the intersection manager, who computes decisions based on a local control policy. Alternatively, constraints may regard the min/max speed profile to hold, as in the case of~\cite{Kowshik2011804}: there, an hybrid architecture for intelligent intersections mixes decentralised computations (as regards collision avoidance) and centralised decision making (for coordinating vehicle traversing the intersection) ultimately letting vehicles decide whether to cross or not in the assigned time frame.

In general terms, all the above approaches ensures safety and starvation by giving every vehicles the possibility to cross the intersection. Most importantly, simulations show that such approaches dramatically reduce the waiting time for vehicles with respect to traditional approaches based on stop signs or traffic lights~\cite{Dresner2008591}, because: \emph{(a)} the occupancy of the intersection is maximised and \emph{(b)} vehicles from different directions can cross the intersection without waiting, provided they are not in direct collision---i.e., they occupy different portion of the intersection, or occupy the same portion at different times.
 
A problem of centralised approaches is that they require the presence of an infrastructural element (the intersection manager) and the capability of vehicles to interact with it. Thus, they can hardly be applicable in the wild. A recent proposal~\cite{8482420} suggests the possibility (in absence of any infrastructure, but only exploiting V2V communication) to be engaged in a leader election algorithm, to elect a transient leader vehicle in charge for a predefined amount of time. The leader, elected by considering factors such as the distance to the intersection, the traveling speed, etc., temporarily acts as intersection manager and (other than acquiring its own right-of-way) elaborates which other vehicles can cross the intersection safely. 

\subsubsection{Negotiation} 

In centralised approaches, vehicles have no word over the policies enforced by the intersection manager. Negotiation-based approach, instead, enables vehicles to actively participate in the protocol aimed at establishing in which order vehicles will gain access to the intersection. 

Such protocol, given the competitive nature of the problem, can take the form of an \emph{auction}. In approaching the intersection, vehicles contact the intersection manager by placing a ``bid'', that is, by making an offer to ``buy'' the portions of the intersections they require for crossing, for the time required to cross. The value of the bid expresses  the urgency of the vehicle in crossing, it is autonomously set by each individual vehicle according to its own strategy, and it can correspond to some real-world currency or some sort of ``road credits'' assigned to vehicles. The intersection manager collects the bids, gives the right-of-way to the set of vehicles that are in a collision free trajectory and, amongst those that are in collision, to the ones having placed the highest bid. 

Examples of auction-based protocols for intersection management are described in~\cite{Vasirani2012},~\cite{Carlino2013}, and~\cite{Cabri2019}. There, different policies to resolve the auction are analysed, based on different strategies put in place by the vehicles in bidding, as well as different strategies by the intersection manager in establishing the winners. Such strategies can also attempt at incentivising fair bidding while discouraging malicious behaviours. 

In general, auction-based mechanisms (with slight variations depending on the adopted strategies) exhibit performances comparable (at times superior) to that of centralised mechanisms: the waiting time of vehicles is again dramatically lower that that of traditional traffic lights. Safety is ensured provided that vehicles respect the ``rules of the game'', and accept waiting when losing the auction. A problem intrinsic of any auction mechanism concerns liveness, i.e., starvation of vehicles, in that the strategy of vehicles in bidding can sometimes prevent others to win auctions, with the risk for them to experience indefinitely long waiting times. 

Depending on whether auctions are performed with virtual or real-world currency, this can also introduce serious fairness problems, that will be discussed in more detail in Subsection~\ref{democracy}.


\subsubsection{Agreement} 

This category increases the degree of autonomy that vehicles retain while being subject of a coordination process, as the coordination protocol is not fixed \emph{a priori}, but arises from the interaction between participants, \emph{dynamically}. In particular, in this category even the \emph{goals} of the coordinating vehicles as well as their strategies to pursue them can change at run-time depending on the agreement that vehicles establish between each other. As far as intersection management is concerned, solving the problem with agreement techniques essentially amounts at giving vehicles incoming to an intersection the possibility of interacting for affecting each others' originals goals (e.g., directions) and priorities. 

An exemplary proposal, specifically conceived for intersection regulation in the context of bimodal traffic (vehicular plus public transport), is discussed in~\cite{DBLP:journals/wias/BalboBP16}. There, agreement between vehicles happens through a repeated communication protocol running between approaching vehicles and buses, with the assistance of an heterogeneous pool of agents representing conflicting goals (such as the need to minimise private vehicles travel time while prioritising public transportation). Depending on both macro and micro scale criteria, in fact, the agents participating in the protocol may decide to prioritise (hence, ultimately giving right-of-way) either private traffic or public transportation, as a result of a conflict resolution meta-protocol.

Another interesting approach models intersection management as a \emph{Distributed Constrained Optimisation Problem} (DCOP)~\cite{vu2018}, that is, interpreting vehicles as a multi-agent system in which each agent has to collaborate with others so as to find an agreement about the best solution possible to the dynamic set of shared constraints they are involved in. This kind of modelling lends itself to a distributed implementation, where each vehicle interacts in a peer-to-peer way with the neighbouring ones to solve a local problem, that is, DCOP limited to those vehicles actually approaching the intersection. For doing so, the involved agents actually undergo a messaging protocol to exchange their current solutions as they try to adjust their individual values to converge to a feasible (hopefully, optimal) solution.

Finally, let us mention the approach we envisioned in~\cite{8114170}, which proposes to adopt \emph{argumentation technologies}~\cite{rahwan2003argumentation}. In particular, we suggest vehicles can engage in open dialogues while approaching the intersection, discussing their beliefs about the best way to approach the intersection, and in case of conflicting needs, arguing with each other about possible ways to avoid that conflict.  During the dialogue, vehicles can change the argumentation strategy, and may evaluate assertions differently based on the dynamic contingencies arising in the meantime. For instance, a vehicle $A$ approaching the intersection in the north-to-south direction can express arguments about its urgency to cross, and can argue that another vehicle $B$ in the east-to-west direction (and thus conflicting with $A$) could/should decide to cross right, as that move would make $B$ reach in any case destination, but would avoid the conflict with $A$. Persuaded by solidity of $A$'s argument, $B$ could eventually decide to turn right.  Although still at the conceptual level, an argumentation-based approach to intersection management show potential for greater flexibility and adaptivity in facing unforeseen situations. In addition, the power of argumentation approaches in the area of autonomous driving is advocated also by other conceptual proposals as a way to solve conflicts and increase trustworthiness and safety of decisions~\cite{fridman2017arguing}.

\subsubsection{Emergent} 

Approaching intersection management by emergence implies giving absolute freedom to vehicles in choosing how to cross intersections, with the only constraints of acting in a safe way and avoiding starvation. To this end, one can let vehicles either: \emph{(a)} play a selfish game where each agent attempts to maximise its expected utility in crossing despite other agents' needs and goals~\cite{nisan2007algorithmic}; or \emph{(b)} be engaged in an implicit, self-organising coordination scheme, where each vehicles responds in a reactive way to the actions of the other agents, according to some sort of ``natural laws'' enforced in the intersection ``ecosystem''~\cite{pheromoncoord-aamas2002}.    In both cases, coordination does not consider an explicit  agreement about what to do.

Emergent approaches of this kind are indeed already at work in the real world. The way (human-driven) vehicles and motorbikes cross unsignalled intersections in many over-congested African and Asian cities is based on drivers' trying to guess each other actions while making subtle movements strategically aimed at affecting those of nearby vehicles. For instance, in~\cite{Mandiau2008121}, the authors interpret the intersection crossing problem using \emph{game theory}, that is, modelling each vehicle as the player of a game involving other approaching vehicles, each playing its own game, thus each having different payoffs and utility models---that is, essentially, each player is unaware of the formalisation of the game others are playing. The proposed approach investigates how to build decision matrices in such a way that minimal information can be assumed by agents while still being able to find a solution for their own game---that is, a safe way to cross the intersection in the lowest possible time. Alternatively, it is possible to model the collective behaviour of vehicles at intersections in terms of a self-organised collective movement, similar to that of flocking birds~\cite{toner1998flocks}.

However, beside the scientific interest that modelling the behaviour of vehicles at intersections in game-theoretic or self-organising ways can have, actually deploying autonomous vehicles that cross intersections by relying on such approaches seems hardly feasible. In fact, delivering guarantees about safety and liveness may be prohibitively difficult or impractical in the general case of emergent approaches to coordination, as these approaches often exploit stochastic decision making and partial, local information. Indeed, in the cited real-world examples of unsignalled intersections, crashes are frequent, and so it is impossible to predict the time-to-cross.

\subsection{Smart parking}

Smart parking is a \emph{resource-oriented} coordination problem, which is inherently competitive in the case of individual vehicles (or vehicles belonging to different companies), whereas collaborative in the case of vehicles part of a company fleet.

In general, to enable smart parking solutions for autonomous vehicles (and for human-driven vehicles as well), it is necessary to augment parking spaces by deploying sensors (e.g., cameras or presence sensors) to detect free parking slots and inform either some centralised control center or directly vehicles/drivers (depending on the adopted solution) about their location. Alternatively, it would also be possible to exploit vehicles themselves as mobile sensors, detecting and informing about available parking spots as they drive. Also, suitable actuator devices may be needed in the transitory phase during which autonomous vehicles would co-exist with human-driven ones, so as, for instance, to signal human drivers that a parking slot has been reserved although not yet physically occupied. We forward the interested read to~\cite{revathi2012smart} for a survey of such technologies.

\subsubsection{Centralised} 

A centralised solution to smart parking in the presence of autonomous vehicles would simply let vehicles communicate their requests (where they need to park and when) to a central authority (i.e., a brokerage service owned by the municipality or, for fleets, some service managed by the fleet company). The central authority takes care of reserving available parking slots to the requesting vehicles, by assigning vehicles to slots according to its own policy (e.g., maximisation of resource utilisation, minimisation of vehicles delay), and then inform each of the vehicles about the parking slot assigned to them. 

It is worth emphasising that a whole lot of solutions to smart parking are currently limited to provide information about free parking spots to human drivers, that then can select the one to reserve and receive guidance accordingly. Such a sort of systems are not of interest for the present paper as no coordination happens at all, even in matching supply and demand, as human intervention is mandatory---see~\cite{Polycarpou2013,7895130} for comprehensive surveys.

Nevertheless, examples of fully or partially automated systems matching, respectively, vehicles and drivers to a given parking lot without the need for human intervention can be found. For instance,~\cite{s141222372} proposes a Cloud-based architecture where a recommendation algorithm based on the Map/Reduce technique is exploited for suggesting the ``best'' car parking lots to users (receiving the recommendation through a mobile app). The matching between a user (hence, vehicle) and a parking lot depends on factors such as friends' car parking suggestions, own preference profile, and users' parking history, hence the match found by the platform has the goal ox maximising each user satisfaction.
A similar approach is taken in~\cite{RePEc:ems:eureri:12467}, where a centralised Cloud infrastructure is still exploited to gather information, disseminate routing paths, and search for vehicle-lot matchings, but the optimal parking place is defined as the most optimal distribution of vehicles over the available parking places---from the perspective of the municipality, for instance. Each user specifies his point of destination, the maximum distance the parking place is allowed to be from the destination, and the maximum fee he is willing to pay for the parking place: these factors are then weighed against the collective interest of all other users to determine the optimal parking place. Moreover, Ant Colony Optimisation~\cite{Dorigo2019} is exploited to provide dynamical navigation paths to vehicles, taking into account the real-time traffic density as monitored by smart lampposts scattered across the road network.

\subsubsection{Negotiation} 

A negotiation approach would let vehicles participate to a coordination protocol where each of them has the right to advance proposals, reject others' ones, suggest alternatives, and the like, according to an individual negotiation strategy (for instance meant at minimising cost or travel distance). In other words, each vehicle may actively participate to the decision making process of coordination.

In~\cite{CHOU2008805} for instance, a decentralised implementation of an \emph{auction} in a multi-agent systems setting is described, where each vehicle undertakes a \emph{two-stage negotiation} with a broker agent responsible of a (set of) parking slot(s): first, it bargains to reserve a parking slot, then for setting the price to pay. Three are the criteria adopted by the broker to match parking demand with supply: parking fees, distance to the parking spot and to the destination, and the broker's own booking and reservation policies (as stemming, e.g., from the owner of the parking lot). In the paper, a few negotiation strategies are also compared: bid contracting, simple contracting, and sequential request with best-first acceptance.

A similar approach is taken in~\cite{Geng20131129}, but leveraging on a centralised implementation where a single entity (actually composed by three logical modules) decides the allocations of demanding vehicles to parking slots. The approach is in the negotiation category as the demanding vehicles have the opportunity to reject an allocation received, and to ask for another one by supplying a set of constraints, over which the two parties may negotiate until an acceptable solution is found. In particular, the allocation process is interpreted as a sequence of mixed-integer linear programming problems solved at specific time points, continuously.

The approach described in~\cite{Ayala201243}, instead, proposes a dynamic pricing scheme adjusted through a \emph{market-oriented optimisation} algorithm, so as to let the parking slot manager influence the way in which the available parking slots are occupied by competing vehicles. The authors validate through simulations that the proposed pricing algorithm is able to steer the system towards load balancing the available resources.

In~\cite{dinapoli2014} an Iterated Contract Net protocol is set up between a parking manager agent and a user agent having conflicting goals: the former tries to achieve fair distribution of occupied slots while avoiding over-crowding the city centre, whereas the latter only wants the closest and most convenient spot. The parking manager makes offers for available parking slots to the user agent, which accepts or refuses them, until convergence to a compromise between user satisfaction and public good is found.

Finally, in~\cite{alves2019experimentation} both collaborative and competitive approaches are evaluated to compare their performance w.r.t.\ paid price and distance to parking spot: Contract Net is used a collaborative approach, while English (lower price first, then raise) and Dutch (opposite) auctions are exploited as competitive ones. Along the distance dimension, all approaches show similar results, whereas along the price metric the Dutch auction is the worst, whereas the other two are comparable.

\subsubsection{Agreement} 

The only example of agreement approaches applied to smart parking we found in the literature is the argumentation-based approach described in~\cite{munoz2010developing}. However, the manuscript lacks an evaluation section, hence no conclusions about feasibility and efficacy of the approach are possible.

The easiest explanation of this under exploration could be that there is not much to agree upon when vehicles are looking for parking: they have the same individual goal, they have no interest in collaborating, hence the target of a potential agreement appears uncertain. Also, possibly, approaches such as DCOP my be overly complex (hence, computationally unsustainable) for the assignment problem that parking allocation essentially is. Argumentation-based approaches may provide some beneficial effects, especially in terms of fairness of parking slots assignment, by letting vehicles debate over their urgency or precedence and let an arbiter decide upon the dispute. For instance, a vehicle may cooperate to accept farther parking slots as its passengers must meet a deadline for an important meeting, whether another one may instead value more a closer spot as its passenger has difficulty in walking.

\subsubsection{Emergent} 

Approaches leveraging coordination arising by emergence from local interactions amongst vehicles are not so common in the context of smart parking, possibly because they are best suited to complex systems where many individual agents need to frequently interact so as to achieve a common goal without centralised control---whereas smart parking often concerns either a single vehicle to be assigned a spot, or a few vehicles competing for one.

Nevertheless,~\cite{Ayala2011299} for instance, proposes to exploit a distributed implementation of a \emph{game theoretic} approach where competing vehicles are modelled as selfish agents striving to obtain parking slots. This setting is dealt with under the assumption of complete knowledge of the game by each player (payoff known), and that of incomplete knowledge instead. Furthermore, a centralised implementation is also presented, where a single authority makes parking choices for travellers and assigns each to a specific parking slot. In both cases, the objective of the game theoretic solution is to optimise social welfare.

\subsection{Ride sharing}

As pointed out in Section~\ref{taxonomy}, ride sharing is a sort of umbrella term for many highly related sub-problems, such as carpooling, and presents differencesst depending on whether private individual vehicles or company-owned fleets are considered, making the problem competitive or cooperative, respectively. 

As anticipated, for taxis and similar services like Uber, most approaches are centralised: a single site collects mobility requests and, based on the availability of vehicles/drives to satisfy them, proposes a match based on its own internal policies. We do not exclude that in the future, in the presence of autonomous vehicles, it will be possible to rely on direct V2V and vehicle-to-user communications for ride sharing or to satisfy more general classes of mobility needs. However, even if, most likely, the approaches will still consider the presence of centralised sites to collect requests and availabilities, the key differences with respect to current approaches will be that: \emph{(a)} they will be able to exploit more detailed information (e.g., about positions of vehicles, traffic situations, etc.) \emph{(b)} they will make it possible to decentralise the decision making process the leads to matches. 

\subsubsection{Centralised} 

Centralised approaches to ride sharing concern solving a constrained optimisation problem aimed at matching supply and demand while complying to time-based and route-based constraints (e.g.\ departure and arrival times, maximum deviation from optimal route, etc.). We forward the reader to~\cite{Agatz2012295} and~\cite{Furuhata201328} for two extensive surveys on these kinds of approaches, also outlining the advantages in efficiency that the availability of detailed information can bring. 

Let us just exemplify two specific proposals for centralised decision making in ride sharing. The work described in~\cite{Herbawi:2012:GIH:2330163.2330219} exploit a \emph{genetic algorithm} to both propose initial matchings between supply (vehicles and drivers) and demand (``free riders''), constrained by pickup and delivery times as well as maximum distance travelled and capacity occupied, and later adjust those matchings ``on-line'' as new riders want to dynamically join. The approach proposed in~\cite{kleiner2011mechanism}, although still centralised in decision making, goes somewhat in the direction of negotiation: the idea is to let passengers bid for drivers by interacting with an individual broker with complete knowledge of the system (drivers availability, departure and arrival times, routes, as well as passengers preferences). However, the interacting parties cannot adjust bids or refuse offers, or anyhow negotiate with each other, or with the broker, to find a better individual solution---hence there is little autonomy left to vehicles (or, drivers). 

As for all centralised approaches to negotiation, the distinction between collaboration and competition, and thus the distinction between private vehicles and fleets of company vehicles is blurred. In fact, by putting themselves in the hand of a central decision maker, even individual vehicles can be interpreted as partss of a fleet. 

\subsubsection{Negotiation} 

A solution to the ride sharing problem based on negotiation techniques, e.g.\ auctions or Contract Net, is possibly the most suitable besides the centralised one, as it is quite natural to model the problem as a set of interacting agents collectively engaging in a joint planning process, and actively pursuing convergence through negotiation, indeed. The benefits that a negotiation-based approach may have are increased flexibility and scalability, as each ``ride sharing group'' may independently (concurrently) find an agreement about the parameters of the joint trip, such as pickup/delivery locations, departure/arrival times, etc.

In the specific context of \emph{carpooling}, that is sharing a vehicle amongst private individuals,~\cite{doi:10.3141/2542-11} presents a multi-agent based simulation framework where \emph{negotiation} between ``carpooling agents'' is at the core of the coordination approach used to match supply and demand, as well as to arrange trips. In particular, the authors assume existence of a ``carpooling social network'' where people with similar home-work and work-home trips are clustered, so as to navigate the graph in search for potential matchings. When a group of drivers sharing similar routes is formed, as soon as requested by one of the participants negotiation takes place at two steps: driver and vehicle selection, and departure time setting. This process may repeat even during trips or while an agreement has been already settled, for instance because members leave the group (or new ones join) or change preferences (such as pickup times, pickup and drop-off order, trip start times, etc.).
It is worth noting that the focus of the paper is on forming carpool groups, disregarding analysis the competition that may arise between different vehicles to maximise their own utility, e.g., monetary gain.

In~\cite{10.1007/978-3-319-33509-4_32} instead the focus is on \emph{fleet management}, namely task allocation and redeployment of vehicles owned by a set of organisations: the former problem amounts at determining which vehicle should serve a given task, whereas the latter consists in relocating vehicles so that new tasks can be better dealt with (faster and/or with lower costs). In particular, the authors exploit an \emph{auction-based} mechanism to assign tasks to vehicles, inspired by Bertsekas' auction algorithm~\cite{Bertsekas1988}. It is worth emphasising that the auction mechanism employed is structured in such a way that the overall system gain is maximised, not the individual vehicle gain. For instance, in the taxi fleet scenario described, vehicles are assigned to passengers by collecting all the requests in a given time window, and then distributing requests to taxis so as to maximise utilisation, not individual passengers' utility function.

\subsubsection{Agreement} 

We could not find any agreement approach in the surveyed literature on ride sharing. However, here the lack of DCOP-based solutions is perhaps surprising, as the problem of assigning vehicles to routes is often times interpreted as an assignment problem which could be handled by DCOP, especially in the case of privately owned vehicles---in the case of the fleet of a company, instead, centralised or negotiation-based solutions seems more practical.

As far as argumentation is concerned, benefits may be gained again especially in the case of privately owned vehicles: each vehicle may argue about an unplanned detour, or a delay in delivery time, and others may attack or support its arguments towards settlement of the debate. Assuming to have the possibility of registering passengers and drivers' preferences, which already is taken into account in some approaches, letting ride sharing undergo an argumentation-based coordination process seems a path worth exploring.

\subsubsection{Emergent} 

Solving ride sharing with an emergent approach would amount at letting vehicles figure out an arrangement for serving ride sharing requests in complete autonomy, by interacting with each other without trying to impose any constraints about what do and when, but solely abiding to some ``rules of the game'' either resembling their own convenience/goal (as for game theory) or somehow ``hard-coded'' in the system (such as for nature-inspired approaches).

The \emph{game theoretic} model proposed in~\cite{Banerjee20151837} for the more specific problem of carpooling is an example of emergent approach: by careful designing incentives to cooperate, in the form of social credit, the authors obtain a distributed coordination process where each participant to the carpooling solves (repeatedly) a stochastic game with the objective to maximise its own utility (hence, social reward). Due to a careful integration of this distributed coordination protocol matching supply and demand with the reward mechanism, the authors achieve the desirable effect of incentivising cooperation while letting individuals maximise their own payoff.

Also the work presented in~\cite{7942006} starts from a \emph{game theoretic} treatment of the problem, but models it in a very different way, drawing inspiration from multi-agent planning theory: in particular, the  solution described allows participants to engage in a collective planning process in terms of a \emph{bargaining game} allowing each driver to seal the deal using its preferred strategy. Once drivers arrive at their decision, a cooperative protocol takes charge of mitigating the conflicts possibly arisen: in particular, every player discloses its strategy to others, so as to let conflicting players start to interact to resolve conflict.

\subsection{Ramp merging}

Ramp merging (and the similar lane changing) is a task-oriented coordination problem that, although inherently competitive, has to be addressed in a cooperative way. 

In most cases, approaches to this coordination problem take as reference a \emph{platoon}-based setting, that is, a scenario in which each of the on-ramp (or source lane) vehicles are considered individually as they merge into the existing platoon of vehicles already travelling in the highway (or destination lane) as an ensemble---and, similarly, vehicles willing to leave the highway are individually considered as splitting ones. However, the problem can be faced also in the absence of a platoon-based setting. 

Whether considering a platoon-based setting or not, the general benefits that a coordination approach to ramp merging may deliver are similar to those for intersection management: stop-and-go behaviour and the need to slow down (both for the merging vehicle and for the one already occupying a lane) may be drastically reduced.

\subsubsection{Centralised} 

Centralised approaches to ramp merging (and lane changing) are adopted in the context of a platoon setting. The leader of the platoon takes care of all the decision making involved about the actions that the other vehicles of the platoon and the entering vehicle have to take in order to facilitate merging/splitting manoeuvres. Vehicles however can also exploit their own sensing and V2V communications in order to calibrate the actions. 

Approaches of this kind are described in~\cite{Awal20131468} and~\cite{Rios-Torres2017780}. Also, most of the many approaches surveyed in~\cite{Rios-Torres20171066} adopt a similar endeavour, by applying either optimisation techniques or model predictive control strategies (as in~\cite{CAO201598}) which leave vehicles no degrees of freedom w.r.t.\ the outcome of the coordination process.

\subsubsection{Negotiation} 

In the absence of an existing leader, that is when vehicles travel on a lane outside of a platoon, a vehicle wishing to enter a lane from a ramp, or in need to change lane, has to somehow make sure to: \emph{(a)} avoid collisions with the vehicles on the entering lane and \emph{(b)} enter the lane smoothly, that is avoiding forcing the vehicles on the entering lane to abruptly reduce speed. This may require some sort of negotiation of relative actions and speeds between the entering vehicle and the other vehicles already on the lane in its proximities. 

In the approach described in~\cite{Amoozadeh2015110}, the actions of the vehicles involved in a ramp merging scenario  are decided via peer-to-peer V2V negotiation protocol, in which the entering vehicle proposes actions to the vehicles already on the target lane, which have the right to refuse and make alternative proposal. A very similar approach is described in~\cite{7945011}. In high-traffic conditions, though, such approaches are at risk of starvation, if the vehicles are not willing or are not in the conditions to accept the proposal of the entering vehicles. 

To avoid starvation, negotiation approaches based on incentive mechanisms for vehicles on the entering lane, such as auctions where merging vehicles offer to pay some kind of currency to enter can be conceptually envisioned. However, we haven't found any relevant example of this kind in the literature.

\subsubsection{Agreement} 

In a ramp merging (or lane changing) scenario it is quite difficult to imagine engineering efficient agreement approache.

A notable exception is described in~\cite{ramasamy2018systems}. There a coordination protocol between 2 or 3 vehicles (but easily extendable to more) is proposed where vehicles coordinate to find the best accomplishment for a lane change operation (essentially the same problem faced in ramp merging scenarios). In particular, vehicles may assess the quality of their moves and exploit such a measure to motivate their proposal to take action (or coordinate action), and also may substantiate the rejection of a proposal during negotiation by communicating the state of affairs leading to such decision. This clearly can be assimilated to a coordination approach based an \emph{argumentation}.

\subsubsection{Emergent} 

Concerning emergent approaches to foster coordination, some game theoretic models can be spotted in the surveyed literature, but are exploited in the context of centralised approaches where a single individual (i.e., the leader of the platoon) tells participant what to do.

A rather unique exception is described in~\cite{Wu20161482}, which proposes a \emph{game-theoretic} approach were two vehicles are considered at a time to play a 3-stages sequential game aimed at letting them agree on the best solution to merge. In the proposed setting two more measures are taken by the authors to find better solutions to the two-players game: communication between vehicles is allowed as a means to negotiate actions (admissible moves), and also multi-agent learning is exploited so as to let vehicles improve their own utility matrix as games are played.

As far as self-organised approaches are concerned, these can be potentially exploited in the context of leaving and entering a platoon by taking inspiration from the behaviour of flocking birds~\cite{toner1998flocks}, as described in the following subsection.

\subsection{Platooning}

Platooning is the cooperative problem of coordinating manoeuvres of a fleet of vehicles so that they travel altogether as a single entity. 

\subsubsection{Centralised} 

Centralised approaches to platooning in the literature typically involve a preliminary phase of electing a platoon leader, which is then in charge of communicating to the other vehicles the speed profile to which they must abide~\cite{Michaud2006437}. Clearly, in need of speed variation or other road contingencies, the leader could communicate these changes to the other vehicles of the platoon, which adapt accordingly~\cite{Amoozadeh2015110}. 

\subsubsection{Negotiation and agreement} 

Negotiation- and agreement-based approaches to platooning have not been found in the literature. Whether it depends on unsuitability of the approach or lack of exploration, one reason may be that there is little to negotiate or agree upon during platoon formation or disbanding: one may claim that a potential subject of negotiation/agreement could be the speed profile to keep, but usually platoon formation exactly serves the purpose of traveling safely at the maximum allowed speed---hence, simply, the head vehicle sets the pace. Along this line, a path worth exploring could be that of negotiating the goal of the platoon, for instance, pursuing maximum speed or most fuel-efficient driving behaviour; nevertheless, to date this appears to be unexplored.

\subsubsection{Emergent} 

Here, the platoon formation does not involve the election of a leader, and groups of vehicles can simply engage in a platoon by having each vehicle dynamically adapt their speed to those of nearby vehicles, exploiting their sensors to do so, and without any explicit agreement~\cite{7056505}. Such an approach can be considered mimicking the self-organising behaviour of those many species of social animals that tend to move in formation~\cite{toner1998flocks}, and in particular those of flocking birds, which do so also for the sake of maximising efficiency of flight (a key goal for vehicle platooning as well). In this context, also the acts of entering and leaving a platoon (that is, the problems of ramp merging and lane changing) can be attacked with such a kind of self-organising approaches.

It is worth mentioning that similar self-organising solutions are also applied to Unmanned Aerial Vehicles, to let them travel in formation despite disruption~\cite{7387990}.
\subsection{Traffic flow optimisation}
\label{optimisation}

Traffic flow optimisation involves directing traffic across a road network (e.g., at an urban scale) in order to balance overall exploitation of the road infrastructure, minimising travel times, and avoiding traffic jams. As discussed in Subsection~\ref{flow}, today's modern approaches attempt at influencing vehicles by means of adapting traffic lights schedule~\cite{Adacher2014992,6799768} and digital signages. In the presence of autonomous vehicles, the approach can be completely different. 

\subsubsection{Centralised} 
\label{flowcentralised}

Centralised solutions to traffic flow optimisation in the presence of autonomous vehicles will naturally be based on similar algorithmic techniques as that of current systems: sensorial infrastructures monitoring the traffic situation, to inform a centralised traffic management systems in charge of applying traffic flow/density forecasting techniques to anticipate congestions and jams~\cite{vlahogianni2014short}, and to elaborating plans avoiding them. 

The actuation phase, however, will be different: instead of acting on traffic lights (which, as we envision, will mostly disappear) and digital signages, vehicles can directly receive guidance from the centralised traffic management system~\cite{myr2003real} instructing vehicles on which routes to avoid. However, providing personalised individual guidance to each vehicle seems impractical, especially in large-scale scenario. Therefore, most likely, instructions will be sent collectively to groups of vehicles on a geographical basis (i.e., instructions to change route sent to all or a given fraction of the vehicles that find themselves in a road where a traffic jams is predicted to occur). 

An alternative option to actuate plans could be that of acting at the level of traffic intersections, as it is done today with traffic lights, but acting at the level of the policies enforced to coordinate intersections~\cite{hausknecht2011}, i.e., dynamically adapting in a coordinated way the parameters adopted in the different intersections, if not dynamically changing the coordination approach adopted in different intersections itself. 

\subsubsection{Negotiation and Agreement} 

Negotiation and agreement approaches to traffic flow optimisation at the level of individual vehicles can hardly apply, indeed we did not found any examples about in the literature. In fact, given the scale of the problem involving a large number of vehicles, elaborating solutions based on distributed negotiations and discussions amongst vehicles  seems impractical and would likely require long delays before reaching a solution. 

However, to overcome the problem, one can think at the possibility of selecting a subset of representative vehicles (e.g., elected as leader or randomly selected) in different regions of the road network. Such vehicles, each accounting for the needs of the region it represents, can then negotiate/agree with each other towards the identification of suitable actions aimed at overall traffic optimisation. A proposal of this kind is described in~\cite{vasirani2011artificial}, where specialised software agents (not necessarily vehicles) are assigned portions of the road infrastructure and are put at negotiating with each other. Indeed, similar approaches to negotiations and agreements are common practice in cities in order define urban mobility plans: multiple stakeholders are typically involved in the decision process, such as representatives of residential neighbourhoods, managers of shopping centres, public transport managers, etc. The aim is to reach a common agreement about the goals and the strategies to enforce with the urban mobility plan. 

\subsubsection{Emergent} 

Emergent approaches to traffic flow optimisation, although not making it possible to control the behaviour of individual vehicles, can guarantee the achievement of specific properties at the global, collective level. For this reason, in particular self-organised ones, can indeed by a viable solution for traffic flow optimisation at the urban scale.

Solutions in the literature mostly take inspiration from the natural world, and from the way the movement of individuals in systems can naturally self-distribute in an environment simply based on indirectly sensing the presence of other individuals. With this regard, in~\cite{1193933,CamurriMZ06}, we have proposed vehicles spreading of virtual computational fields across the road networks, and their dynamic aggregation, so as to express sort of ``density gradients''. Vehicles can then move towards their destination trying to get the direction with decreasing intensity of such field, that is, by moving away from congested roads. Simulation results show that this enables improving the balance of the traffic over the road infrastructure, and the traffic flow consequently. A similar approach is proposed in~\cite{5409626} that takes instead inspiration from biology: virtual computational pheromones are deposited by vehicles as they travel. Such pheromones tend to repel other vehicles, with the results of limiting the concentration of vehicles and increasing distribution. 

Besides the scenario of autonomous vehicles, it is also worth outlining that similar nature-inspired approaches have been proposed (e.g., in~\cite{zou2019self,ho2019improved}) as a means to properly orchestrate the schedule of traffic lights, and to orchestrate the movements of robots aimed at mapping and exploring complex environments~\cite{shen2004hormone}.

\subsection{Summarising discussion}

Given the diversity of coordination problems we have overviewed and the different approaches that can be adopted to attack them, it is quite difficult to draw some ultimate conclusions about what approach better suits which problem. Nevertheless, some generals considerations can be made:

\begin{itemize}

\item Centralised approaches offer the best guarantees in terms of safety and liveness, thanks to the strict control over the actions of individual vehicles. Also, they make it possible for e.g., urban traffic managers or managers of vehicles' fleets, to impose their own policies on traffic behaviour. However, for large-scale problems (i.e., traffic flow optimisation or parking allocation at the urban scale), centralised approachess may fall short, as the duty of controlling individual vehicles can become overwhelming (other than less relevant). As an additional drawback of centralised approaches, they leave no room to vehicles for personalising the driving experience, whether it's due to the contingent need for a vehicle to cross an intersection quickly, or to find a parking in a specific zone, or to catch a timely ride. 

\item Decentralised negotiation approaches, by enabling agents to express their needs and by having the coordination process account for them, can help somehow accommodating individual needs, but at risk of losing fair treatment of all vehicles during the coordination process, to the point of failing in liveness (e.g., inducing starvation of vehicles). From the safety viewpoint, it can be ensured by a properly designed negotiation protocol and the strict adherence of vehicles to it.

\item Agreement appears promising in solving conflicts between vehicles, in finding suitable coordination schemes between them (e.g., at intersections or at the more global urban level), and even at better evaluating conditions which may be critical for safety~\cite{fridman2017arguing}. However, such approaches are definitely the less explored ones in the literature, probably because argumentation theory so far is well-studied from a theoretical standpoint in areas such as sociology and multi-agent systems, but still lacks effective practical development frameworks and software engineering methods.

\item Emergent approaches appear not suitable for those coordination problems where there is the need of guaranteeing accurate and predictable behaviour at the level of individual vehicles, such as in crossing intersections or in ramp merging. They can be more properly applied when the overall global-scale behaviour of the system, rather that of individual vehicles, is the key focus of the coordination problem. This is the case of traffic flow optimisation, or parking (if looked at from the perspective of traffic managers) or even ride sharing. An exception to this rule concerns vehicles' platooning, where emergent self-organising solutions assimilable to those expressed in social animals (and already extensively applied to swarms of UAVs) are indeed applied.

\end{itemize}
In any case, to give more solid ground to the above general considerations, and to make it possible to define more specific guidelines, more studies are necessary. In particular, there is need of systematic, comparative, and quantitative analysis of the dependability (safety and liveness) and performances (quality measures) of the different coordination approaches applied to the different coordination problems in a variety of different traffic conditions. 

With this regard, we emphasise that the scientific literature on concurrency control, distributed systems, social systems, and multi-agent systems has plenty of solutions and prior experience to share. For instance, negotiation-based coordination approaches have been extensively studied in multi-agent systems literature, likewise agreement and emergent ones in the area of complex and social systems.

\section{Horizontal Challenges}
\label{challenges}

Let us now introduce some ``horizontal'' challenges related to the coordination of autonomous vehicles, i.e., general issues that apply to all the problems and approaches discussed so far. The coordination approaches that we have discussed can be hardly realised and deployed in the real-world in a dependable and acceptable way without also identifying suitable solutions to these additional challenges

\subsection{Dynamic switching of coordination scheme}

In previous section, for the different coordination problems, we have presented different approaches based on a different degree of autonomy left to vehicles. However, a single solution can hardly handle all possible situations that can occur. 

For instance, in the case of intersection management, a solution based on direct negotiation or argumentation between vehicles can be very effective in low traffic situations, when the number of vehicles involved in such negotiations is quite low, thus a collective outcome can be reached quickly. Instead, in the case of congested traffic situations, with a large number of vehicles involved, reaching a shared agreement can be harder and induce notable overhead and delay in communications. Also, in the case of auctions, it can induce inflationary effects on the bids. Hence, in these situations, it would be better to rely on a more centralised solution based on an intersection manager~\cite{Carlino2013}. Or, as discussed in~\cite{hausknecht2011} it can be necessary for different intersections to dynamically adapt the coordination approach in order to support traffic flow optimisation. As another example, a centralised parking scheme that works well let the city governance control the distribution of parked vehicles, may fall short in the presence of a high-number of vehicles, by inducing notable delays in parking. In this case it is better to switch to an approach that lets individual vehicles negotiate for parking slots according to their own preferences~\cite{dinapoli2014}. 

The above issue suggests the need to dynamically switch from one type of coordination solution to another, upon changing conditions. This implies dynamically varying the level of autonomy of vehicles: for instance, in heavy-traffic conditions, an intersection manager could be instantiated reclaiming autonomy of decisions from vehicles and taking charge of decisions itself. 

In the area of robotics and multi-agent systems the theme of ``coordination with adjustable autonomy''~\cite{mostafa2017} (sometimes referred to as ``flexible autonomy'', also~\cite{gerber1999flexible}) has been extensively discussed, either referring to the fact that, at times, a human actor may wish to reclaim autonomy in decisions from agents or robots to take care of decisions herself~\cite{scerri2001}, or to the fact that (as in our scenario) specific conditions may require to dynamic switch coordination scheme~\cite{van2008}. In the real world, and in the context of safety-critical situations such as the coordination of autonomous vehicles, though, designing and realising such dynamic switch can be conceptually and technically very hard: who decides when to switch? How can a group of cars agree on the if and when of switching? Who takes care of coordinating the switching act? Who instantiates/activates the additional components (e.g., a traffic intersection manager)? We do not have solutions ready to use, but certainly the vast amount of literature on adjustable autonomy can suggest useful research directions ~\cite{mostafa2017}.

\subsection{Coordination at the ``System of Systems'' level}

In previous section, we addressed the issue of coordination mostly at the level of individual systems (a single intersection, a single entering lane, or the cars of a single car sharing company), and discussed how such problems can be dealt with in isolation. However, thinking at  a more global level, it is clear that the coordination actions in one system may impact on other systems. That is, coordination issues arise also at the level of ``systems of systems''. For example, we all know that queues at an intersection can induce queues at nearby intersections, or that similarly a slow down in a motorway due to an intense flux of traffic in an entering lane can quickly propagate backwards to impact previous entering lanes. As other examples, when a car or ride sharing company starts having problems in satisfying all requests, this may impact on the requests arriving at competing companies, suggesting inter-company coordination and agreement in serving request peaks. 

To add further complexity, one should also think that municipalities (as discussed in Subsection~\ref{optimisation}) might wish to impose specific policies over traffic management, for instance to reduce overall pollution and/or noise, or to avoid excessive crowding in specific parts of a city. Similarly, motorway managers may wish to impose limitations on the flux of entering vehicles to avoid over-crowding and slowing down. 

Both the inter-related effects of individual coordination acts and the need to respect global level policies will necessarily imply that the solutions and the policies adopted to solve an individual coordination problem cannot be designed without accounting for the systemic impact of such solutions and policies. Or, in other words, that the level of individual coordination must be coupled with a meta co-coordination level, in which an agreement at the global level is reached on how to act (i.e., according to which policies and constraints) at the local level. In the area of autonomous vehicles, a few work exist that handles such systemic problems. For instance, in~\cite{Vasirani2012}, it is analysed how global coordination of intersections can be achieved by trying to affect, at the local level, the choices of individual vehicles. A similar analysis is presented in~\cite{hausknecht2011}. 
In different fields (logistics, energy management, teamwork, robotics, and multi-agent systems) a variety of mechanisms have been proposed for coordination in large-scale systems of systems (e.g., hierarchical mechanisms, market-based, self-organising)~\cite{Thompson1991,OmiciniZ16}. Such mechanisms can be a source of inspiration for the field of autonomous vehicles as well.

\subsection{Ethics of coordination acts}

The safety-critical nature of autonomous vehicles implies the necessity of devising solutions to handle unexpected situations. Indeed, all the presented coordination schemes consider dynamic context-aware decisions that account for unexpected situations and tune movements of vehicles accordingly, e.g., slowing down in the presence of pedestrians, halting to give road to an ambulance, or accounting for ``byzantine'' vehicles that gets temporarily disconnected and are unable to participate in the coordination process. However, in the presence of critical situations that can cause damage to cars and road infrastructures, or even injuries and death to passengers and pedestrians around, deciding how to act becomes not simply a problem of algorithmic nature, but a problem of ethical nature. 

Concerning individual autonomous vehicles, the most representative examples of ethical decisions that they should face could be expressed as specific instantiations of the well-known trolley dilemma\footnote{The trolley dilemma is expressed as follows: a trolley is advancing on a railway and is going to kill three persons that are in its trajectory. You are seeing the scene, by a rail switch, and are in a position to operate the switch so as to deviate the trolley towards a direction where it would kill a single person. What should you do? Nothing, and let three people die. Or switch, thus saving three people but making yourself directly responsible of the death of the other persons that would have otherwise survived?}~\cite{Foote1978}. For example, consider an autonomous vehicle that abruptly finds in its trajectory a child, and has not enough time and space to stop the vehicle. The only alternative to save the child would be to turn the car and have it crash against a wall, thus killing its passenger(s).  What should the vehicle do? Simply nothing, and kill the child, or deviate towards the wall killing the passenger(s)? What is the most ethical decision?  Besides the fact that, as a passenger, I would hardly accept to ride a vehicle that can decide to deliberately kill myself, programming the behaviour of autonomous vehicles to handle such situations requires a shared code of ethics, backed up by a proper regulatory framework dictating the responsibilities shared by manufacturers, programmers, passengers, owners, etc. Unfortunately,  legislation comprehensively ruling autonomous vehicles is yet to be defined, and this is further complicated by the fact that no universally shared code of ethics can be defined to handle such situations (as from the recent ``moral machine experiment''~\cite{Awad2018}).

In the case of coordinating autonomous vehicles, a variety of different instances of the trolley dilemma can be conceived, i.e., a variety of critical situations that can occur during the process of coordination. Most of these can be expressed in terms of cars that are in collision with each other, with the need of deciding which one must deviate from the planned trajectory towards an unsafe one, or in terms of groups of cars that finds themselves in a position to cause damage or injury, and have to decide who will take actual responsibility for this. The additional complexity that these types of critical decisions express over the simpler single vehicle case arises from the need of agreeing on a common ethical choice amongst different vehicles, and from the fact that different vehicles (e.g., coming from different nations) may have different ethical guidelines embedded.   

\subsection{Road neutrality and traffic democracy}
\label{democracy}

Another problem that might emerge in future scenarios of autonomous vehicles concerns usage of traffic infrastructures and of mobility services, and this can be again a sort of ethical, or better "democracy'' problem. As of today, while driving, we are already used to pay for the usage of infrastructures, such as parking slots, bridges, motorways. Similarly, we pay for mobility services such as buses and taxis. Common characteristics of all of these cases are that \emph{(a)} the payments are based on fixed, known, and universal fees; \emph{(b)} the service received is neutral. 

With regard to the former point, whether I drive my car on Saturday night or on a busy Tuesday morning, I will always pay the same known fee on the Milano-Venice motorway. Also, I will pay the same fee whether I drive a Ferrari or a small city car. Mobility services can have different fees at different hours and day, but again based on public and clear fee schemes.  With regard to the latter point, whether I have a Ferrari or a city car, I will receive exactly the same quality of service from a parking slot, or while driving on a motorway, and there is no way for a richer person to have a better service by paying more. That is, the use of mobility services and infrastructures express neutrality over the users. With the advent of autonomous and coordinating vehicles, things can dramatically change.

If vehicles can dynamically request the usage of infrastructures and pay it automatically, it may become possible for the manager of such infrastructures to impose dynamic pricing mechanisms~\cite{Chen2018}, based on the current demand load.  Something similar is already happening with regard to mobility services (consider, e.g., the dynamic pricing mechanisms of Uber and of airlines companies), but at least we are always given a choice to accept the proposed price before buying the services. Applying such dynamic schemes to cars that are moving around implies that a vehicle, while starting its ride, may have no a priori idea on how much it will eventually cost~\cite{Brent2018}. Also, one can consider that the possibility for cars to dynamically pay while on-transit, and to dynamically track exactly which infrastructure a vehicle is using, opens up the way for imposing fees on the usage of infrastructures other than parking slots and motorways. For instance, as in the already presented auction-based approaches to intersection management, it could become possible to impose payments for crossing busy intersections, with fees varying in dependence of traffic and time-to-wait, or for using a specific lane in motorways with fees depending on the average speed on that lane.

The mechanisms of dynamic payment, could also enable a model in which passengers can decide to pay more to get better services, breaking the current neutrality of road infrastructures. Consider the mechanisms of crossing intersections based on virtual auctions that we have introduced in Section~\ref{survey}. In the future, such mechanisms could become based on a real auction with real money, with the consequence that richer vehicles will always bid higher and buy priority in crossing the intersection, while poorer vehicles will risk starvation. Similar issues can arise in motorways w.r.t.\ priority lanes. 

The above issues are not of an algorithmic or technical nature, but calls for the definition of suitable regulations to avoid mobility becoming a privilege.

\subsection{Mixing human drivers and autonomous cars}

Most of the considerations we have made so far in this paper assumed that all the cars involved in coordination are of a fully autonomous nature, or at least that (during the coordination act) they act and interact autonomously with each other without human intervention. Such assumption can match a not-so-near future when we can expect that human-driven cars will no longer exist or when, for safety and efficiency reasons, it will be forbidden for humans to drive but in specific controlled situations (the same as today, for instance, it is forbidden to ride a horse in motorways and high-speed roads). However, there will be a rather long transition phase in which our streets will be populated by a mixture of fully autonomous cars, partially autonomous ones, and traditional human-driven cars. Such a scenario clearly challenges the possibility of relying on the surveyed coordination schemes, unless one devises out dependable means to involve human-driven cars (that is, their human drivers) in the process of coordination. 

In the area of robotics and autonomous systems, the issue of coordinating mixed teams of robots and humans has been extensively analysed, and many possible means of interactions have been devised, e.g., for motion planning and task partitioning in rescue or war operations~\cite{Thrun2004,Goodrich2008}. These scenarios have several characteristics in common with vehicle coordination, but assume the existence of means for robots and humans to communicate with each other. In a vehicle coordination scenario, this would imply human-driven cars to be necessarily equipped with suitable hardware to send/receive information to/from nearby cars and road infrastructure, and the presence of friendly interfaces (e.g., based on natural language, wearables, augmented reality, and/or properly modified infrastructural traffic lights~\cite{dresner2007sharing}) for humans to receive information while driving. Fortunately, such features can be easily integrated in existing cars, and are cheap enough that we can expect to become legally compulsory.

Concerning the coordination schemes to be applied in a group of mixed human-driven and autonomous vehicles, these can be designed according to the solutions discussed in Section~\ref{survey}, properly modified and tuned to account for the possible inaccuracy of actions by human-driven cars. For instance, a human-driven car should not be asked to cross an intersection with one centimetre accuracy in position and a millisecond accuracy in time. In any case, amongst the many surveyed schemes, those based on argumentation~\cite{8114170} appear the most natural ones for humans to be involved in, given their dialogical nature.

The issue of mixing human-driven and fully autonomous cars, though, is a very complex one, and would require a full survey on its own.

\section{Conclusions}\label{conclusions}

For autonomous vehicles to start circulating in our streets and cities, it will be necessary to identify solutions for coordinating their relative movements in order to let them circulate safely and without conflicts and crashes. In this article, we firstly introduced a taxonomy for the different classes of coordination problems of autonomous vehicles, and overviewed the individual coordination problems according to their classes. Then we tried to frame the key solutions that can be identified for such coordination problems, showing that the distinguishing characteristics of the different solutions concern the level of autonomy in decision making assigned to vehicles during the coordination process.

Although there are still many open challenges to address before autonomous vehicles can safely hit the road, we hope our attempt at framing some key concepts and potential solutions can help the research in this area, one that promises to be a flourishing source of plenty of challenging and fascinating research problems to face.


\begin{acks}
The authors would like to thank the Italian MIUR PRIN 2017 Project ``Fluidware'' for supporting this work.
\end{acks}

\bibliographystyle{ACM-Reference-Format}
\bibliography{MCZ-traffic-survey-2019}

\end{document}

%% file: problems.tex
%
\begin{table}[!t]

\caption{Coordination problems for connected autonomous vehicles: bi-dimensional taxonomy.}
\setlength\extrarowheight{10pt}
\ltab{problems}
\begin{center}
\begin{tabular}{| c || c | c |}
\hline
~ & \textbf{Resource-oriented} & \textbf{Task-oriented}\\[10pt]
\hline
\hline
\textbf{Competitive} & \makecell{intersection manag.\ \\ parking (private)} & \makecell{ride sharing (private) \\ ramp merging}\\[10pt]
\hline
\textbf{Collaborative} & \makecell{parking (fleet) \\ traffic flow opt.} & \makecell{ride sharing (fleet) \\ platooning}\\[10pt]
\hline
\end{tabular}
\end{center}
\end{table}
%

%% file: mapping.tex
%
\begin{table*}[!t]
\footnotesize
\renewcommand{\arraystretch}{1.5}

\caption{Traffic problems mapped to coordination problems, in terms of their core, defining elements. Emphasis is used to highlight the resource / \emph{task} oriented nature of problems, as per \xt{problems}.}
\ltab{mapping}
\begin{center}
\setlength\extrarowheight{5pt}
\begin{tabular}{| c || c | c | c | c | c |}
\hline
~ & \textbf{Resource} / \textbf{\emph{Task}} & \textbf{Safety} & \textbf{Liveness}& \textbf{Quality Measures}\\[5pt]
\hline
\hline
\makecell{\textbf{intersection}\\ \textbf{management}} & intersection & no collision & no starvation &  \makecell{avg.\ / max.\ / cumulative \\ delay, travel time, \\ throughput, queue length}\\[5pt]
\hline
\textbf{smart parking} & parking slot & no overbooking & no exclusions & \makecell{total distance,\\ resource usage,\\ user satisfaction,\\ avg.\ time to park,\\ wandering ratio}\\[5pt]
\hline
\textbf{ride sharing} & \makecell{\emph{serve mobility}\\\emph{requests}} & \makecell{no overlapping roles\\ no disconnected trips} & no exclusions & \makecell{travel time,\\\ vehicle-miles,\\ avg.\ no.\ matching,\\ \% saved cars,\\ carpools lifespan}\\[5pt]
\hline
\textbf{ramp merging} & \emph{merge / split} & no collision & no starvation & \makecell{avg.\ merging time / rate,\\ waiting time,\\ avg.\ velocity,\\ avg.\ throughput} \\[5pt]
\hline
\textbf{platooning} & \emph{behave as one} & no collision & no fragmentation & --- \\[5pt]
\hline
\makecell{\textbf{traffic flow}\\\textbf{optimisation}} & road infrastructure & no traffic jams & fair scheduling & \makecell{travel time} \\[5pt]
\hline
\end{tabular}
\end{center}
\end{table*}
%

%% file: approaches.tex
\begin{table}[!b]

\caption{Criteria for definition of and attribution to autonomy classes of the coordination approaches for connected autonomous vehicles. The adjectives in each cell refer to the vehicles participating to the coordination process.}
\setlength\extrarowheight{10pt}
\ltab{approaches}
\begin{center}
\begin{tabular}{| c || c | c | c | c |}
\hline
~ & \textbf{Strategy} & \textbf{Protocol} & \textbf{Role} & \textbf{Degree of Freedom}\\[2pt]
\hline
\hline
\textbf{Centralised} & external (coordinator) & fixed & passive & none\\[2pt]
\hline
\textbf{Negotiation} & individual & dynamic & active & admissible moves\\[2pt]
\hline
\textbf{Agreement} & individual & dynamic & active & admissible moves, goals\\[2pt]
\hline
\textbf{Emergent} & individual & none & active & full\\[2pt]
\hline
\end{tabular}
\end{center}
\end{table}

%% file: survey.tex
\begin{table*}[!t]
\setlength\extrarowheight{10pt}
\caption{Problems vs.\ coordination approaches, categorised and ordered according to increasing \emph{autonomy} allowed for vehicles (left to right).}
\ltab{survey}
\begin{center}
\small
\begin{tabular}{ c || c | c | c | c }
\hline
~ & \textbf{Centralised} & \textbf{Negotiation} & \textbf{Agreement} & \textbf{Emergent}\\[5pt]
\hline
\hline
\makecell{\textbf{Intersection}\\\textbf{management}} & \cite{Wu201565,6121907,6338827,Dresner2008591,Kowshik2011804,8482420} & \cite{Vasirani2012,Carlino2013,Cabri2019} & \cite{DBLP:journals/wias/BalboBP16,vu2018,8114170} & \cite{Mandiau2008121}\\[5pt]
\hline
\textbf{Smart parking} & \cite{s141222372,RePEc:ems:eureri:12467} & \cite{CHOU2008805,Geng20131129,Ayala201243,dinapoli2014,alves2019experimentation} & \cite{munoz2010developing} & \cite{Ayala2011299}\\[5pt]
\hline
\textbf{Ride sharing} & \cite{Agatz2012295,Furuhata201328,Herbawi:2012:GIH:2330163.2330219,kleiner2011mechanism} & \cite{doi:10.3141/2542-11,10.1007/978-3-319-33509-4_32} & --- & \cite{Banerjee20151837,7942006}\\[5pt]
\hline
\textbf{Ramp merging} & \cite{Awal20131468,Rios-Torres2017780,Rios-Torres20171066,CAO201598} & \cite{Amoozadeh2015110,7945011} & \cite{ramasamy2018systems} & \cite{Wu20161482}\\[5pt]
\hline
\textbf{Platooning} & \cite{Michaud2006437,Amoozadeh2015110} & --- & --- & \makecell{\cite{7056505}\\ \cite{7387990} (UAVs)} \\[5pt]
\hline
\makecell{\textbf{Traffic flow}\\\textbf{optimisation}} & \makecell{\cite{myr2003real,hausknecht2011}\\ \cite{Adacher2014992,6799768,vlahogianni2014short} (forecast)} & \cite{vasirani2011artificial} & --- & \makecell{\cite{1193933,5409626,CamurriMZ06}\\ \cite{zou2019self,ho2019improved} (traffic lights)\\ \cite{shen2004hormone} (robots)}\\[5pt]
\hline
\end{tabular}
\end{center}
\end{table*}